\documentclass[11pt]{article}
\usepackage{graphicx}
\graphicspath{{C:/Users/xinwei/Documents/6EPMapping/gunterPaper/}}

\columnwidth\textwidth

\def\al{\alpha}

\def\ro{\varrho}

\def\eh{{\scriptstyle{1\over 2}}}
\def\d{\partial}
\def\=d{\,{\buildrel\rm def\over =}\,}

\def\te{\vartheta}
\def\B{\Bigl}

\def\diag{{\rm diag}}
\def\e{{\rm em}}
\def\o{{\rm obs}}

\begin{document}

\title{CMB in non-standard cosmology: A first look }
\author{G\"unter Scharf
\footnote{e-mail: scharf@physik.uzh.ch}
\\ Physics Institute, University of Z\"urich, 
\\ Winterthurerstr. 190 , CH-8057 Z\"urich, Switzerland}

\date{}

\maketitle\vskip 3cm

\begin{abstract} 

We study CMB  in the nonstandard background cosmology recently investigated. Using the previously calculated first order metric perturbations we discuss the Sachs-Wolfe and the integrated Sachs-Wolfe effects. We show how small-multipole CMB data can be used to determine the matter density of the Universe.


\end{abstract}

\newpage

\section{Introduction}

The cosmological standard FLRW model is a high-density model. On the contrary nonstandard cosmology is a low-density model where only the few percent of visible matter contribute to the energy density [1]. As far as observations can decide between the two, the magnitude-redshift data are excellently reproduced by both models (see next section). The next step is the analysis of CMB which consists of two parts:
(i) Understanding the early universe so far that one gets initial conditions at the time of last scattering, for: (ii) Propagating CMB in the cosmic gravitational field from last scattering to the present. The standard model solves both problems successfully. For the nonstandard model we solve (ii) in this paper, that is our first look. Problem (i) is much harder because the nonstandard early Universe is quite different from the standard one.

The paper is organized as follows. In the next two sections we review the nonstandard background and its first order perturbations.
 In sect.4 we calculate the CMB temperature anisotropy by applying a formula of Tomita [2]. In the discussion we point out how data of small multipoles $l$ can be used to determine the matter density in the Universe. This is complementary to standard cosmology where small $l$-values are neglected because of foreground effects. As a preparation we discuss the transformation of CMB data to the cosmic rest frame in the appendix.

\section{Nonstandard background and its Hubble diagram}

The nonstandard background is defined by the line element
$$ds^2=dt^2-X(t)^2dr^2-R(t)^2(d\te^2+\sin^2\te d\phi^2)\eqno(2.1)$$
in comoving spherical coordinates. The time dependence of the metric functions $R(t)$ and $X(t)$ is given in parametric form by [1] 
$$R(t)=T_L\sin^2w,\eqno(2.2)$$
$$X(t)=\cot w\eqno(2.3)$$ 
where the comoving time is equal to
$$t=T_L(w-\sin w\cos w).\eqno(2.4)$$
Here $T_L$ determines the lifetime of the Universe. From (2.2) it follows
$$\dot R(t)=2T_L\sin w\cos w{dw\over dt}.\eqno(2.5)$$
and (2.4) yields
$${dt\over dw}=2T_L\sin^2w.\eqno(2.6)$$
Then we obtain
$$\dot R(t)={\cos w\over\sin w}=\cot w=X.\eqno(2.7)$$

For the physical discussion we need the radial null geodesics given by the wave vector $k^\mu=(1/X,-1/X^2,0,0)$. Then the redshift is given by
$$1+z={X_\o\over X_\e}={\cot w_\o\over\cot w_\e}\eqno(2.8)$$
where $\e$ and $\o$ refer to the time of emission and observation, respectively. The Big Bang corresponds to $z=\infty$, that means $w_\e=\pi/2$, and $\pi/2<w_\o<\pi$, because $z$ (2.8) must be positive. From
$$dz={\cot w_\o\over\cot^2w}{dw\over\sin^2w}={\cot w_\o\over\cos^2w}{dt\over 2T_L\sin^2w}\eqno(2.9)$$
we identify the Hubble constant
$${dz\over dt}\B\vert_{z=0}=-H_0={(1+\cot^2w_\o)^2\over 2T_L\cot w_\o}.\eqno(2.10)$$

To calculate the radial distance we integrate
$${dr\over dz}={dr\over dt}{dt\over dz}={2T_L\cot ^2w\over X\cot w_\o}={2T_L(1+z)^3\over [(1+z)^2+\cot^2w_\o]^2}$$
from $z=0$ to $z>0$. With the new variable of integration $x=1/(1+z)$ we get
$$r(z)=2T_L\int\limits_{1/(1+z)}^1{dx\over x(1+\cot^2w_\o x^2)^2}.\eqno(2.11)$$
We introduce the parameter
$$\al={1\over\vert\cot w_\o\vert}\eqno(2.12)$$ 
which is also equal to the local light speed $c_0=dr/dt=1/\vert X\vert$. Using the Hubble constant (2.10) we finally obtain
$$r(z)={c_0\over H_0}(1+\al^2)^2\int\limits_{1/(1+z)}^1{dx\over x(\al^2+x^2)^2}.\eqno(2.13)$$
The luminosity distance is equal to $(1+z)r(z)$. The rational integral in (2.11) is elementary so that
$$d_L(z)={c_0\over H_0}(1+z){(1+\al^2)^2\over 2\al^2}\B[{1\over 1+\al^2}-{(1+z)^2\over 1+\al^2(1+z)^2}+$$
$$+{1\over\al^2}\log{1+\al^2(1+z)^2\over 1+\al^2}\B].\eqno(2.14)$$
The magnitude $m(z)$ is defined by
$$m(z)=5\log_{10}d_L+M+25\eqno(2.15)$$
where $M$ is the absolute magnitude of the supernova standard candle. In the Hubble diagram one plots the distance modulus
$$\mu(z)=m(z)-M.\eqno(2.16)$$

The parameter $\al^2$ will now be determined by the Hubble data.
The measured Hubble diagram is nicely represented by the standard FLRW luminosity distance
$$\tilde d_L(z)={c_0(1+z)\over H_0}\int\limits_1^{1+z}{dx\over\sqrt{\Omega_Mx^3+\Omega_\Lambda}}.\eqno(2.17)$$
From the type Ia supernovae observations one has obtained the following parameter values $\Omega_M=0.27$, $\Omega_\Lambda=0.73$, $H_0=72$ km/(s Mpc). This is the best fit in [3]. In the table the corresponding distance modulus $\tilde\mu(z)$ is listed in the second column. We have taken the value
$\tilde\mu(1)=44.08$ at $z=1$ as a measured value and have determined the free parameter $\al^2$ in (2.14) such that this value is reproduced. The result is
$$\al^2=6.71.\eqno(2.18)$$
With this value the entire Hubble diagram until $z=10$ is excellently represented by (2.14-16) as can be seen in the third column of the table. The last two columns show the look-back times in the standard and nonstandard models [1].

$$\vbox{\halign{#&#&#&#&#&#&#&#\cr
$z$\qquad\quad&$\tilde\mu(z)({\rm mag})\qquad$&$\mu(z)({\rm mag})\qquad$&$\tilde t(z)(10^9Y)\qquad\qquad$&$t(z)(10^9Y)\qquad$\cr
\noalign{\hrule\vskip 0.2 cm}
0.01&33.12&33.12&0.1349&0.1348\cr
0.02&34.64&34.64&0.2678&0.2676\cr
0.03&35.53&35.54&0.3990&0.3985\cr
0.04&36.17&36.18&0.5283&0.5275\cr
0.05&36.67&36.68&0.6558&0.6546\cr
0.06&37.08&37.09&0.7816&0.7799\cr
0.07&37.43&37.44&0.9057&0.9034\cr
0.08&37.74&37.75&1.0281&1.025\cr
0.09&38.01&38.02&1.1488&1.145\cr
0.1&38.25&38.26&1.2679&1.263\cr
0.2&39.89&39.91&2.3756&2.360\cr
0.3&40.89&40.91&3.3443&3.317\cr
0.4&41.62&41.64&4.1969&4.158\cr
0.5&42.20&42.22&4.9489&4.903\cr
0.6&42.69&42.71&5.6145&5.565\cr
0.7&43.10&43.12&6.2054&6.157\cr
0.8&43.46&43.48&6.7317&6.690\cr
0.9&43.79&43.80&7.2020&7.171\cr
1.0&44.08&44.08&7.0236&7.608\cr
2.0&46.05&45.96&10.181&10.45\cr
3.0&47.22&47.03&11.318&11.90\cr
4.0&48.05&47.78&11.928&12.79\cr
5.0&48.70&48.36&12.300&13.38\cr
6.0&49.22&48.82&12.541&13.80\cr
7.0&49.67&49.21&12.711&14.12\cr
8.0&50.05&49.55&12.836&14.37\cr
9.0&50.38&49.84&12.93&14.57\cr
10.0&50.68&50.10&13.033&14.73\cr
}}$$

\section{The metric perturbations}

In the following it is convenient to use the variable
$$x={1\over X(t)}\eqno(3.1)$$
so that
$$R=T_L{x^2\over x^2+1}.\eqno(3.2)$$
This $x$ is directly related to the redshift (2.8)
$$x=\al(1+z)\eqno(3.3)$$
and $\al$ is the parameter (2.12). Due to (2.10) $T_L$ is proportional to the Hubble time:
$$T_L={(\al^2+1)^2\over 2H_0\al^3}.\eqno(3.4)$$
The components of the inverse nonstandard metric are equal to
$$g^{\mu\nu}=\diag(1,-1/X^2,-1/R^2,-1/R^2\sin^2\te).\eqno(3.5)$$
The non-vanishing Christoffel symbols are given by
$$\Gamma^0_{11}=X\dot X=-{\dot x\over x^3}\eqno(3.6)$$
$$\Gamma^0_{22}=R\dot R={2T_L^2x^3\over (x^2+1)^3}\dot x\eqno(3.7)$$
$$\Gamma^0_{33}={2T_L^2x^3\over (x²+1)^3}\dot x\sin^2\te\eqno(3.8)$$
$$\Gamma^1_{01}={\dot X\over X}=-{\dot x\over x}\eqno(3.9)$$
$$\Gamma^2_{02}=\Gamma_{03}^3={\dot R\over R}={2\dot x\over x(x^2+1)^2}\eqno(3.10)$$
$$\Gamma^2_{33}=-\sin\te\cos\te,\quad\Gamma^3_{23}=\cot\te.\eqno(3.11)$$

In the 2+2 formalism of Gerlach and Singupta [4] the linear perturbations of the metric  are classified into even-parity or polar perturbations and odd-parity or axial perturbations. Only the former are relevant in the following and are given by ([4], equ.(2.4))
$$h_{\mu\nu}dx^\mu dx^\nu=h_{AB}(x^C)Y(\te,\phi)dx^Adx^B+h_A(x^C)Y,_a(dx^Adx^a+dx^adx^A)+$$
$$+R^2[KY(\te,\phi)\gamma_{ab}+G(Y,_{a\vert b}+{l\over 2}(l+1)Y(\te,\phi)\gamma_{ab})]dx^adx^b.\eqno(3.12)$$
Here capital Latin indices refer to $x^0=t$ and $x^1=r$ and small Latin indices refer to the angles $\te$ and $\phi$, $Y(\te,\phi)$ are the spherical harmonics where the indices $l$ and $m=-l,\ldots +l$ are always omitted because they are decoupled on the spherically symmetric background. The comma always means partial derivatives and the vertical bar denotes covariant derivatives on $M^2$ (spanned by $x^C$, $C=0,1$) and on the unit two-sphere (spanned by $\te,\phi$ with metric tensor $\gamma_{ab}$), respectively [4]. Furthermore we have shown in [1] that only the following components are different from zero
$$h_{00}=-{1\over x^2}H_2,\quad h_{01}=H_1$$
$$h_{11}=-x^2H_2,\quad R^2K=T_L^2\B({x^2\over x^2+1}\B)^2K\eqno(3.13)$$
This has the important consequence that $h_{\mu\nu}$ coincides with the gauge invariant perturbations $k_{\mu\nu}$ which are needed for the CMB calculations [2]. After Fourier transform in the radial coordinate
$$\hat f(x,q)=(2\pi)^{-1/2}\int f(x,r)e^{iqr}dr\eqno(3.14)$$
the three functions $\hat H_1$, $\hat H_2$ and $\hat K$ are solutions of the differential equations [1]
$$(x^5+x^3)\hat K'=-(x^4+3x^2)\hat K-2x^2\hat H_2-l(l+1)(x^2+1)\hat H_4$$
$$(x^7+2x^5+x^3)\hat H'_2=-(x^6+4x^4+3x^2)\hat K+(2x^6+6x^4+4x^2)\hat H_2-[Q^2x^6+l(l+1)(x^4+2x^2+1)]\hat H_4\eqno(3.15)$$
$$(x^5+2x^3+x)\hat H'_4=2x^4(\hat K+\hat H_2)+2(x^4+2x^2+1)\hat H_4.$$
Here the prime denotes derivative with respect to $x$ and we have introduced the quantities
$$\hat H_4={\hat H_1\over iqT_L}\eqno(3.16)$$
$$Q^2=2q^2T_L^2\eqno(3.17)$$
where $q$ is the wave number.

To calculate the metric perturbations one has to integrate the system (3.15) from last scattering $z=1090$ to present time $z=0$, that means from $x=1091\al$ to $x=\al=2.59$. It is important to notice that (i) there is no $m$-dependence in (3.15) and (ii) the dependence on $l$ appears in the form $l(l+1)$. This has two consequences: (i) If the initial condition for (3.15) were axially symmetric then CMB must be axisymmetric around the direction of the dipole anisotropy. (ii) One should study CMB observables as functions of $l(l+1)$ instead of simply $l$. We shall return to this point in Sect.5. Regarding (i) we do not know whether the degree of axisymmetry of CMB has been analyzed. We investigate this in the Appendix by transforming CMB data to the cosmic rest frame. 

\section{Null geodesics and the CMB anisotropy}

We assume that the observer of CMB sits at the origin $r=0$ in the cosmic rest frame. That means its comoving coordinates coincide with the cosmic rest frame. Measurements on Earth must be corrected for the motion of the Earth. Then CMB radiation arrives on radial null geodesics. Let
$$K^\mu={dx^\mu\over d\lambda}\eqno(4.1)$$
be the wave vector of radiation, $\lambda$ is an affine parameter. The radial wave vector has only two non-vanishing components $A=0,1$ satisfying
$$K^AK_A=(K^0)^2-X^2(K^1)^2=0\eqno(4.2)$$
and it fulfills the geodesic equation
$$K^A_{\vert B}K^B=(K^A,_B+\Gamma^A_{BC}K^C)K^B=0.\eqno(4.3)$$
By (3.6) the zeroth component $A=0$ reads
$$K^0,_0K^0+K^0,_1K^1+x\dot X(K^1)^2=0.$$
Here we insert (4.2)
$$K^1=\pm {K^0\over X}\eqno(4.4)$$
obtaining
$$K^0,_0\pm {1\over X}K^0,_1+{\dot X\over X}K^0=0.\eqno(4.5)$$
The $A=1$ component gives the same equation. Since $X(t)$ does not depend on $r$, the general solution is
$$K^0={C\over X},\quad K^1=\pm {C\over x^2}.\eqno(4.6)$$

For an inward light ray we must choose the minus sign in (4.6). Then the affine parameter satisfies
$$d\lambda={1\over K^0}dx^0+{1\over k^1}dx^1={C\over x}dt-{C\over x^2}dr.$$.
Using $dr=-x dt$ we have
$$d\lambda={2C\over x}dt.\eqno(4.7)$$
Below we shall choose the normalization $C=1/2$ which yields
$$K^A={x\over 2}(1,-x).\eqno(4.8)$$
Instead of the affine parameter $\lambda$ we shall use the physical variable $x$ (3.3). The relation between the two is given by the auxiliary variable $w$ 
$$x=\vert\tan w\vert.\eqno(4.9)$$
On the other hand $w$ is connected with the comoving time $t$ by (2.4)
$$t=T_L(w-\sin w\cos w).\eqno(4.10)$$
This implies
$$dt=2T_L\sin^2w dw\eqno(4.11)$$
Here we substitute $\sin w$ by $\tan w$ which gives $x$ (4.9) and insert $dw=dx/(x^2+1)$. Then we obtain
$$d\lambda={2T_Lx\over(x^2+1)^2}dx.\eqno(4.12)$$
Consequently $\lambda$ can be identified with $R(t)$ (3.2). For the Christoffel symbols (3.6) we also need
$$\dot x={1\over 2T_L\sin^2w\cos^2w}={(x^2+1)^2\over 2T_Lx^2}.\eqno(4.13)$$

We now consider the measured CMB temperature anisotropy
$$\Delta T=\sum_{lm}a_{lm}Y_l^m(\te,\phi).\eqno(4.14)$$
Tomita [2] has derived an elegant formula for $a_{lm}$ as an integral over the light path from last scattering $\lambda_e$ to the present $\lambda_f$:
$$a_{lm}=-{1\over 2}\int\limits_{\lambda_e}^{\lambda_f}{d\lambda\over K^0}\B[g_{0D}(k_{DB\vert C}+k_{DC\vert B}-k_{BC\vert D})-$$
$$-k_{BC}{K^0_{\vert 1}+\Gamma^0_{D1}K^D\over K_1}\B]K^BK^C.\eqno(4.15)$$
By (3.13) the gauge invariant metric perturbations are given by
$$k_{00}=h_{00}=-{1\over x^2}H_2(x,r)\eqno(4.16)$$
$$k_{01}=h_{01}={iQ\over\sqrt{2}}H_4(x,r)\eqno(4.17)$$
$$k_{11}=h_{11}=-x^2H_2(x,r).\eqno(4.18)$$
The radial derivative $\d_1=\d/\d r$ produces a factor
$$-iq=-i{Q\over\sqrt{2}T_L}\eqno(4.19)$$
in the Fourier transform (3.14) and the temporal derivative is calculates as
$$\d_0={1\over 2T_L\sin^2w}{\d\over\d w}={(x^2+1)^2\over 2T_Lx^2}{\d\over\d x}.\eqno(4.20)$$
Then for the covariant derivatives of the perturbations we get
$$k_{00\vert 0}=\d_0k_{00}=-{(x^2+1)^2\over 2T_Lx^2}\d_x\B({H_2\over x^2}\B)$$
$$k_{01\vert 1}=\d_1k_{00}+2{\dot x\over x}k_{01}={iQ\over\sqrt{2}T_Lx^2}\B(H_2+{(x^2+1)^2\over x}H_4\B)\eqno(4.21)$$
$$k_{01\vert 0}=\d_0k_{01}+{\dot x\over x}k_{01}={iQ\over\sqrt{2}}{(x^2+1)^2\over 2T_Lx^2}\B(H_4'+{H_4\over x}\B)$$
$$k_{01\vert 1}=\d_1k_{01}+{\dot x\over x}k_{11}+{\dot x\over x^3}k_{00}={Q^2\over 2T_L}H_4-{(x^2+1)^2\over 2T_Lx}H_2\B(1+{1\over x^6}\B)$$
$$k_{11\vert 0}=\d_0k_{11}+2{\dot x\over x}k_{11}=-{(x^2+1)^2\over 2T_L}\B(H_2'+{4\over x}H_2\B)$$
$$k_{11\vert 1}=\d_1k_{11}+2{\dot x\over x^3}k_{01}={iQ\over\sqrt{2}T_L}\B(x^2H_2+{(x^2+1)^2\over x^5}H_4\B).$$
We note that the third function $K$ does not contribute, consequently, a glance to (3.12) shows that the angular dependence is really given by the simple spherical harmonics $Y_l^m(\te,\phi)$ as in (4.14). For simplicity we have omitted the hats for radial Fourier transform.

Now using (4.8) the first term under the integral (4.15) becomes
$${x^2\over 4}k_{00\vert 0}-{x^3\over 4}k_{00\vert 1}+{x^3\over 4}k_{01\vert 0}+{x^4\over 2}k_{01\vert 1}-{x^4\over 4}k_{11\vert 0}=$$
$$=H'_2{(x^2+1)^2\over 8T_L}\B(x^4-{1\over x^2}\B)+H_2{(x^2+1)^2\over 4T_L}x^3-{iQx\over 4\sqrt{2}T_L}H_2+$$
$$+H'_4{(x^2+1)^2\over 8\sqrt{2}T_L}iQx+{H_4\over 4T_L}\B(Q^2x^4-{iQ\over 2\sqrt{2}}(x^2+1)^2\B).\eqno(4.22)$$
Furthermore we have
$$K^0_{\vert 1}+\Gamma⁰_{D1}K^D={\dot x\over x}={(x^2+1)^2\over 2T_Lx^3}$$
which must be divided by $K_1=-g_{11}x^2/2=1/2$ and multiplied by
$$k_{BC}K^BK^C=-{H_2\over 4}(x^6-1)-H_4{iQx^3\over 2\sqrt{2}}.$$
In the final formula we must also include the Fourier transformation, the integral $d\lambda$ is transformed by (4.12) into an integral over $x$. Then we obtain the following formula
$$a_{lm}=(2\pi)^{-1/2}\int\limits_{x_e}^{x_f}dx\int dq\,e^{-iqr}{2x\over(x^2+1)^2}\B\{{\hat H_2\over 4}\B[{iQ\over\sqrt{2}}+(x^2+1)^2\B({x^2\over2}+{1\over x^4}\B)\B]-$$
$$-{\hat H'_2\over 8}(x^2+1)^2\B(x^3-{1\over x^3}\B)+\hat H_4\B[{9\over 8}{iQ\over\sqrt{2}}{(x^2+1)^2\over x}-{Q^2x^3\over 4}\B]-
{\hat H'_4\over 8}{iQ\over\sqrt{2}}(x^2+1)^2\B\}.\eqno(4.23)$$

The radial coordinate $r$ in the exponential must be derived from
$${dr\over dt}=-{1\over X}=-x$$
Using (4.13) we have
$${dr\over dx}=-{2T_Lx^3\over (x^2+1)^2}\eqno(4.24)$$
which after integration yields
$$r=\int\limits_\al^x{2T_Lx'^3\over (x'^2+1)^2}dx'=$$
$$=T_L\B(\log{x^2+1\over \al^2+1}+{1\over x^2+1}-{1\over\al^2²+1}\B).\eqno(4.25)$$

The two terms with $\hat H'_2$ and $\hat H'_4$ in (4.23) can be integrated by parts in $x$. Since $x_e=1091\al$ is very big, the main contribution comes from the term
$$a_{lm}^{\rm SW}=(2\pi)^{-1/2}x_e^4\int\limits_0^\infty dq\cos qr_e{\hat H_2(x_e)\over 2}\eqno(4.26)$$
where we have written the Fourier integral in real form, using the fact that $\hat H_2$ is even in $q$. This contribution is conventionally called the Sachs-Wolfe effect because it only depends on the initial conditions at last scattering $x_e$. The contribution of $\hat H'_4$ is smaller, and all other integral terms give the so-called integrated Sachs-Wolfe effect.

\section{Discussion}

The functions $\hat H_2$, $\hat H_4$ and $\hat K$ appearing in (4.23) have been calculated in [1] as the following power series in $x^{-1}$:
$$\hat H_2=\sigma_l^m(q)\B[{1\over x}\B({12\over 2Q^2+9}-1\B)+{1\over x^3}\B({10l(l+1)+210\over 2Q^2+25}-{6l(l+1)+54\over 2Q^2+9}-2\B)\B].\eqno(5.1)$$
$$\hat H_4=\sigma_l^m(q)\B[-{1\over x}{8\over 2Q^2+9}+{1\over x^3}\B({4l(l+1)+36\over 2Q^2+9}-{4l(l+1)+84\over 2Q^2+25}\B)\B]\eqno(5.2)$$
$$\hat K=\sigma_l^m(q)\B[{1\over x}+{1\over x^3}\B(2-{4l(l+1)+12\over 2Q^2+9}\B)\B]\eqno(5.3)$$
Here $\sigma_l^m(q)$ is the so-called spectral function which specifies some normalization and initial condition for the ordinary differential equations (3.15). These power series must be used as initial conditions for big enough $x$. Then the functions can be calculated for small $x$ by numerical integration of the ordinary differential equations (3.15). The main problem that remains is the determination of the spectral function $\sigma_l^m(q)$. In standard cosmology the corresponding quantity is assumed to be a stochastic variable with a simple covariance like
$$\langle \sigma_l^m(q)\sigma_{l'}^{m'}(q')\rangle=\delta_{ll'}\delta_{mm'}\delta (q-q').\eqno(5.4)$$
This greatly simplifies the problem, but is not very physical because we have only one Universe and one position of observation, so that the expection value in (5.4) has no direct physical foundation.

In non-standard cosmology the situation is completely different. We accept that we live in one particular Universe and near a preferred place which is the center of spherical symmetry of CMB. Then $\sigma_l^m$ is not stochastic. Instead it is a distinguished function which contains information about the initial conditions at the early Universe and the history of the Universe, and about our special place in the Milky Way, and also about the motion of the earth. Then $\sigma_l^m$ cannot be determined by theory alone, it must be derived from other observations. For this purpose the matter density is well suited. By equ. (4.21) of [1] its Fourier transform $\hat\ro (q)$ directly gives the spectral function in (5.1-3):
$$16\pi GR^2\hat\ro_l^m(q)=H_2\B(3+{3\over x^2}+(l-1)(l+2)\B)-K\B[(l-1)(l+2)-2+{Q^2x^6\over (x^2+1)^2}+(x^2+1)\B({1\over 2}+{3\over 2x^2}\B)\B]-$$
$$+H_4\B[{Q^2x^2\over x^2+1}-l(l+1){(x^2+1)^2\over x^4}\B].\eqno(5.5)$$
The power series (5.1-3) can be used for small $l$ only. This is no harm because the surveys as the Sloan Digital Sky Survey give the matter distribution for small redshift only which corresponds to small $l$ values. So in non-standard cosmology the small $l$ region is most interesting where standard cosmology has nothing to say. The comparison of CMB with matter density measurements can help to solve the problem of the dark stuff. In all fairness we see that the more difficult density measurement (5.5) is more valuable than the CMB data (4.23), because the latter is only an integrated quantity. However, for (5.5) the density must be measured in more detail as a function of redshift and angles.
We emphasize that lensing and rotation curves in galaxies test the gravitational fields only ([5], Sect.6.4), {\bf not its sources}. Until now the only real tests of dark matter are the direct searches with underground particle detectors. Then, until now, dark matter does not exist. As a consequence, on galactic scales the gravitational field strongly deviates from Newtonian gravity [5]. But there is no need for modifying Einstein's theory. On the other hand the density (5.5) is intimately connected with the non-standard background (3.1), not as a source but as a first order anisotropic perturbation. The smallness of the baryonic matter density and the small CMB multipole coefficients for $l\le 4$ seem to indicate that this picture is right. Furthermore, the lack of isotropy and homogeneity in the visible matter distribution is another indication for the wrong background of the standard model. But the final answer is open until all observations have been calculated quantitatively.

\appendix

\section{Appendix: Transformation of CMB to the cosmic rest frame}

In the analysis of CMB data this transformation is simply carried out by subtracting the dipole anisotropy. This is a good approximation because the motion of the earth is slow. The exact transformation requires the transformation of spherical harmonics by means of rotation matrices. The temperature anisotropy in galactic coordinates $(\te, \phi)$ is given by
$$\Delta T_l=\sum_{m=-l}^l a_{lm}Y_l^m(\te,\phi).\eqno(A.1)$$
The spherical harmonics are rotated into the cosmic rest frame $(\te',\phi')$ according to
$$Y_l^m(\te,\phi)=\sum_{m'}R_{lm'}^{m}(\al,\beta,\gamma)^+Y_l^{m'}(\te',\phi')\eqno(A.2)$$
Here
$$R_{lm'}^m(\al,\beta,\gamma)=e^{im'\al}d_{lm'}^m(\beta)e^{im\gamma}\eqno(A.3)$$
are the rotation matrices and $(\al,\beta,\gamma)$ the Euler angles of the rotation between the two systems. The $(2l+1)\times (2l+1)$ matrix $d$ is given by [6]
$$d_{lm'}^m(\beta)=\B[{(l+m')!(l-m')!\over(l+m)!(l-m)!}\B]^{1/2}(\cos{\beta\over 2})^{m'+m}(\sin{\beta\over 2})^{m'-m}\times$$
$$\times P_{l-m'}^{(m'-m,m'+m)}(\cos\beta)\eqno(A.4)$$
where $P$ are the Jacobi polynomials. The upper plus in (A.2) denotes the hermitian conjugate rotation matrix which transforms the galactic multipole moments $a_{lm}$ into $a'_{lm'}$ in the cosmic rest frame. By (A.1-2) the multipole moments $a'_{lm'}$ in the cosmic rest frame are given by
$$a'_{lm'}=\sum_m R_{lm'}^{m*}a_{lm}.\eqno(A.5)$$

The WMAP observations [7] (table 6) give multipole moments $\tilde a_{lm}$ for $l=1,2$ and 3.
 But they use real valued harmonics without the usual normalization factor [6] $\sqrt{(2l+1)/4\pi}$. 
As a consequence our moments $a_{lm}$ are obtained from the WMAP values by
$$a_{lm}={1\over\sqrt{2}}(\tilde a_{lm}+i\tilde a_{l-m}.)\eqno(A.6)$$
This formula holds for positive $m$, for negative $m$ we use
$$a_{l-m}=(-)^ma_{lm}^*.\eqno(A.7)$$
WMAP also give the dipole components in galactic rectilinear coordinates (in $\mu$K)
$$(x,y,z)=(-239.3,-2223.\,6,2505.0)\eqno(A.8)$$
The corresponding unit vector is
$$\vec n=(-0.07126,-0.66216,\,0.74596).\eqno(A.9)$$
This gives the direction of the new $z'$-axis. The Euler angle $\beta$ is the angle between the old and the new $z$-axis, hence
$$\cos\beta=n_3=0.74596,\quad\beta=0.728821.\eqno(A.10)$$
The angle $\al$ remains arbitrary which is the freedom of defining the meridian $\phi'=0$ in the new frame. We choose $\al=0$. But $\gamma$ is fixed by the requirement that the dipole has vanishing $x', y'$ components in the cosmic rest frame. 

For the dipole $l=1$ the rotation matrix (A.3-4) is given by
$$R_{m'}^m=\pmatrix{\eh(1+\cos\beta)e^{i\gamma}&-\eh\sqrt{2}\sin\beta e^{i\gamma}&\eh(1-\cos\beta)e^{i\gamma}\crcr
\eh\sqrt{2}\sin\beta&\cos\beta&-\eh\sqrt{2}\sin\beta\cr\eh(1-\cos\beta)e^{-i\gamma}&\eh\sqrt{2}\sin\beta e^{-i\gamma}&\eh(1+\cos\beta)e^{-i\gamma}\cr}
\eqno(A.11)$$ 
By (A.6) the WMAP multipole moments are equal to
$$a_1=-{1\over\sqrt{2}}(239.3+i2223.6)=a_{-1}^*,\quad a_0=2505.0.\eqno(A.12)$$
The vanishing of $a'_{-1}=0=a'_1$ gives the two equations
$$a_1{1\over 2}(1+\cos\beta)e^{i\gamma}+a_0{1\over\sqrt{2}}\sin\beta+a_{-1}{1\over 2}(1-\cos\beta)e^{-i\gamma}=0$$
$$a_{1}{1\over 2}(1+\cos\beta)e^{-i\gamma}-a_0{1\over\sqrt{2}}\sin\beta+a_{-1}{1\over 2}(1-\cos\beta)e^{i\gamma}=0\eqno(A.13)$$
This fixes $\gamma$
$$\gamma=4.8196=276.14^0\eqno(A.14)$$
and yields the transformed $m'=0$ component
$$a'_0=-\sqrt{2}\sin\beta{\rm Re}\B(e^{-i\gamma}a_1\B)+a_0\cos\beta=3358.0.\eqno(A.15)$$
Then the dipole anisotropy is equal to
$$\Delta T_2=a'_0Y_1^0=a'_0\cos\te'.\eqno(A.16)$$
On the other hand the motion of the earth with a velocity $v_1/c=\beta_1$ gives rise to a Lorentz boost which changes the temperature according to [8]
$$T'={T\over \gamma_1(1+\beta_1\cos\te')}$$
where $T=2.725^0$K is the CMB mean temperature and $\gamma_1=(1-\beta_1^2)^{-1/2}$. Expanding this in powers of $\beta_1$ gives an anisotropy.
$$\Delta T=T'-T=T\B[-\beta_1\cos\te'-{\beta_1^2\over 6}+{2\beta_1^2\over 3}P_2(\cos\te')+O(\beta_1^3)\B].\eqno(A.17$$
Comparing this with (A.16) shows that the dipole anisotropy is transformed away by a boost with velocity
$$\beta_1=0.00123,\quad v_1=369 {\rm km/s}\eqno(A.18)$$
in the cosmic rest system. So the latter is obtained from the galactic frame by a rotation plus boost.

The $5\times 5$ rotation matrix for the quadrupole mode $l=2$ is equal to
$$\Biggl({e^{2i\gamma}\over 4}(1+\cos\beta)^2;{e^{2i\gamma}\over 2}\sin\beta(1+\cos\beta);\sqrt{6}{e^{2i\gamma}\over 4}\sin^2\beta;
{e^{2i\gamma}\over 2}\sin\beta(1-\cos\beta);$$ $${e^{2i\gamma}\over 4}(1-\cos\beta)^2$$
$${e^{i\gamma}\over -2}\sin\beta(1+\cos\beta);{e^{i\gamma}\over 2}(2\cos\beta-1)(1+\cos\beta);\sqrt{{3\over 2}}e^{i\gamma}\sin\beta\cos\beta;
{e^{i\gamma}\over 2}(2\cos\beta+1)(1-\cos\beta);$$ $${e^{i\gamma}\over 2}\sin\beta(1-\cos\beta)$$
$${\sqrt{6}\over 4}\sin^2\beta;-\sqrt{{3\over 2}}\sin\beta\cos\beta;{3\over 2}\cos^2\beta-{1\over 2};\sqrt{{3\over2}}\sin\beta\cos\beta;$$
$${\sqrt{6}\over 4}\sin^2\beta$$
$${e^{-i\gamma}\over -2}\sin\beta(1-\cos\beta);{e^{-i\gamma}\over 2}(1-\cos\beta)(1+2\cos\beta);-\sqrt{{3\over 2}}e^{-i\gamma}\sin\beta\cos\beta;
{e^{-i\gamma}\over 2}(2\cos\beta-1)(1+\cos\beta);$$ $${e^{-i\gamma}\over 2}\sin\beta(1+\cos\beta)$$
$${e^{-2i\gamma}\over 4}(1-\cos\beta)^2;{e^{-2i\gamma}\over -2}\sin\beta(1-\cos\beta);\sqrt{6}{e^{-2i\gamma}\over 4}\sin^2\beta;
{e^{-2i\gamma}\over -2}\sin\beta(1+\cos\beta);$$ $${e^{-2i\gamma}\over 4}(1+\cos\beta)^2\Biggr).\eqno(A.19)$$
Here the matrix elements in each row are separated by colons.

WMAP give the following quadrupole moments
$$a_{-2}={1\over\sqrt{2}}(-14.41+i18.80)=a_2^*\quad a_{-1}={1\over\sqrt{2}}(0.05+i4.86)=-a_1^*\quad a_0=11.48.\eqno(A.20)$$
To carry out the rotation this must be multiplied by the columns of (A.19) complex conjugated due to (A.5). We obtain the following final results
$$a'_2=a'_{-2}=14.76+i7.829,\quad a'_{1}=-a^{,*}_{-1}=-1.0222+i7.4777,\quad a'_0=6.621.\eqno(A.21)$$
Here $a'_0$ gets modified by the contribution $P_2(\cos\te')=Y_2^0$ from the Lorentz boost (A.17), we must subtract the quantity
$$T{2\over 3}\beta_1^2=2.748\mu{\rm K}.\eqno(A.22)$$

Finally, the $7\times 7$ rotation matrix for $l=3$ is equal to ($\cos\beta=c$,  $\sin\beta=s$) 
$$\Biggl({e^{3i\gamma}\over 8}(1+c)^3;{e^{3i\gamma}\over -8}\sqrt{6}s(1+c)^2;\sqrt{15}{e^{3i\gamma}\over 8}s^2(1+c);
{e^{3i\gamma}\over -4}\sqrt{5}s^3;$$ $${e^{3i\gamma}\over 8}\sqrt{15}s^2(1-c);{e^{3i\gamma}\over 8}\sqrt{6}s(1-c)^2:{e^{3i\gamma}\over 8}
(1-c)^3$$
$${e^{2i\gamma}\over 8}\sqrt{6}s(1+c)^2;{e^{2i\gamma}\over 4}(3c-2)(1+c)^2;\sqrt{{5\over 2}}{e^{2i\gamma}\over 4}s(1+c)(1-3c);
{e^{2i\gamma}\over 4}\sqrt{30}c(1-c^2);$$ $${e^{2i\gamma}\over 4}\sqrt{{5\over 2}}s(1-c)(3c+1);{e^{2i\gamma}\over 4}(1-c)^2
(3c+2);{e^{2i\gamma}\over 8}\sqrt{6}s(1-c)^2$$
$$\sqrt{15}{e^{i\gamma}\over 8}s^2(c+1);\sqrt{{5\over 2}}{e^{i\gamma}\over 4}s(1+c)(3c-1);{e^{i\gamma}\over 8}(1+c)(15c^2-10c -1);\sqrt{3}{e^{i\gamma}\over 4}s(1-5c^2);$$
$${e^{i\gamma}\over 8}(1-c)(15c^2+10c-1);{\sqrt{5\over 2}}{e^{i\gamma}\over 4}s(1-c)(3c+1);\sqrt{15}{e^{i\gamma}\over 8}s^2(1-c)$$
$${\sqrt{5}\over 4}s^3;{\sqrt{30}\over 4}(1-c^2)c;{\sqrt{3}\over 4}s(5c^2-1);{5\over 2}c^3-{3\over 2}c;$$
$${\sqrt{3}\over 4}s(1-5c^2);{\sqrt{30}\over 4}(1-c^2)c;-{\sqrt{5}\over 4}s^3$$
$$\sqrt{15}{e^{-i\gamma}\over 8}s^2(1-c);\sqrt{{5\over 2}}{e^{-i\gamma}\over -4}s(c-1)(1+3c);{e^{-i\gamma}\over 8}(1-c)
(15c^2+10c-1);\sqrt{3}{e^{-i\gamma}\over 4}s(5c^2-1)$$ 
$${e^{-i\gamma}\over 8}(1+c)(15c^2-10c-1);\sqrt{{5\over 2}}{e^{-i\gamma}\over-4}s(1+c)(1-3c);\sqrt{15}{e^{-i\gamma}\over 8}s^2
(1+c)$$
$$\sqrt{6}{e^{-2i\gamma}\over 8}s(1-c)^2;{e^{-2i\gamma}\over 4}(1-c)^2(3c+2);\sqrt{{5\over 2}}{e^{-2i\gamma}\over -4}s(c-1)(3c
+1);\sqrt{30}{e^{-2i\gamma}\over 4}(1-c^2)c;$$ $$\sqrt{5\over 2}{e^{-2i\gamma}\over -4}s(1+c)(3c-1);{e^{-2i\gamma}\over 4} (1+c)^2(3c-2);\sqrt{6}{e^{-2i\gamma}\over 8}s(1+c)^2$$
$${e^{-3i\gamma}\over 8}(1-c)^3;\sqrt{6}{e^{-3i\gamma}\over 8}s(1-c)^2;\sqrt{15}{e^{-3i\gamma}\over 8}s^2(1-c);\sqrt{5}{e^{-3i\gamma}
\over 4}s^3$$ $${e^{-3i\gamma}\over 8}s^2(1+c);\sqrt{6}{e^{-3i\gamma}\over -8}s(1+c)^2;{e^{-3i\gamma}\over 8}(1+c)^3
\Biggr).\eqno(A.23)$$

The octupole moments in the galactic system have been measured as follows
$$a_{-3}={1\over\sqrt{2}}(11.24+i33.46)=-a_3^*,\quad a_{-2}={1\over\sqrt{2}}(22.03-i0.70)=a_2^*$$
$$a_{-1}={1\over\sqrt{2}}(13.05+i2.45)=-a_1^*,\quad a_0=-5.99.$$
Again to get the value in the cosmic rest frame we have to multiply this by the columns of (A.23) complex conjugated yielding
$$a'_{-3}=-23.57+i8.3142=-a^{,*}_3,\quad a_{-2}=6.4259+i16.0625=a_2^*$$
$$a'_{-1}=-4.5013-i0.7044=-a^{,*}_1,\quad a'_0=-8.353.\eqno(A.24)$$
The additional transformation by the Lorentz boost is smaller than the errors of the measurements. There is no sign of axisymmetry
in these results. Obviously there are strong forground effects. But the same are present in density measurements so that a comparison would give important information.

\end{document}